\documentclass[10pt,twoside]{article}

\usepackage{amssymb}
\usepackage[latin1]{inputenc}
\usepackage{amsmath}
\usepackage{graphicx}
\usepackage{pifont}
\usepackage{psfrag}
\usepackage{enumerate}

\numberwithin{equation}{section}

\begin{document}

\title{Renormalization Group and Grand Unification with 331 Models}
\author{Rodolfo A. Diaz$^{(1)}$\thanks{%
radiazs@unal.edu.co}, D. Gallego$^{(1,2)}$, R. Martinez$^{(1)}$\thanks{%
remartinezm@unal.edu.co} \\
$^{(1)}$Universidad\ Nacional de Colombia, Departamento de Física, Bogotá,
Colombia.\\
$^{(2)}$ The Abdus Salam International Center for Theoretical Physics,
Trieste, Italy.}
\date{}
\maketitle

\begin{abstract}
By making a renormalization group analysis we explore the possibility of
having a 331 model as the only intermediate gauge group between the standard
model and the scale of unification of the three coupling constants. We shall
assume that there is no necessarily a group of grand unification at the
scale of convergence of the couplings. With this scenario, different 331
models and their corresponding supersymmetric versions are considered, and
we find the versions that allow the symmetry breaking described above.
Besides, the allowed interval for the 331 symmetry breaking scale, and the
behavior of the running coupling constants are obtained. It worths saying
that some of the supersymmetric scenarios could be natural frameworks for
split supersymmetry. Finally, we look for possible 331 models with a simple
group at the grand unification scale, that could fit the symmetry breaking
scheme described above.


\end{abstract}

\section{Introduction}

Since the birth of the Standard Model (SM) many attempts have been done to
go beyond it, and solve some of the problems of the model such as the charge
quantization and the unification of the gauge couplings. In some cases the
unification is done by taking a simple group of grand unification, arising
the so called Grand Unification Theories (GUT), where the three interactions
described by SM are treated as only one \cite{Unification}-\cite{Unification3}, the most common
GUT's are $SO(10)$ and $E_{6}$. The first condition for these kind of
theories is an equal value for the three couplings at certain scale of
energy, $M_{U}$. This condition cannot be fulfilled by the simplest grand
unification schemes with the minimal SM particle content and taking the
precision low-energy data. However, the minimal supersymmetric SM can
achieve this scenario for the coupling constants \cite{SUSY}-\cite{SUSY3}. Other
possibilities for unification are the introduction of more degrees of
freedom like fermions and scalar fields that lead the three couplings to
converge at a high energy scale \cite{willenbrock}. Polychromatic extensions
of the SM i.e. $SU(N)_{C}\otimes SU(2)_{L}\otimes U(1)_{1/N}$ have been
considered where the unification of gauge couplings is achieved with $N=7,\ 5
$ and with two, three Higgs doublets, respectively \cite{branco}.
Alternative proposals of unification of quarks and leptons at TeV scale were
considered too \cite{buras}. Finally, another interesting alternative
consists of enlarging the electroweak sector of the SM gauge group such that
the renormalization group equations (RGE) could lead to unification of the
three gauge couplings at certain scale $M_{U}$,$\ $in which there is no
necessarily a group of grand unification at the scale of convergence of the
couplings. In particular, the model based on the $SU(3)_{C}\otimes
SU(3)_{L}\otimes U(1)_{X}$ gauge group (hereafter $331$ models) is an
interesting choice that could address problems like the charge quantization 
\cite{Pires}-\cite{Pires2} and the existence of three families based on cancellation of
anomalies \cite{anomalias}-\cite{anomalias6}.

One way to look for new Physics in 331 models is to check for $tT$
production with $t$ denoting the ordinary top quark and $T$ being an exotic
quark with charge $2/3$ or $4/3$ according to the model considered. In LHC
the production channels would be $pp\rightarrow X^{0}\rightarrow tT$, when
the charge of $T$ is $2/3$, and $pp\rightarrow X^{++}\rightarrow tT$, in the
case in which $T$ has an exotic charge $4/3$. In the first case, in order to
identify the signals, the decay of $t$ is known and identified by the energy
spectrum and angular distribution of the final fermions \cite{Kane et}-\cite{Kane et3}, then
this signal is correlated with the one of an exotic particle with the same
charge of the top, but with a totally different decay in the final state.
The decays of $T$ would be of the form $T\rightarrow X^{0}t\rightarrow \nu Et$%
, $T\rightarrow K^{+}b\rightarrow \nu Eb$, which could de easily
identifyable if exotic leptons have already been produced at LHC. In the
second case in which $T$ has an exotic charge $4/3$, it should be looked for
decays of the type $T\rightarrow X^{++}t$ where $X^{++}$ is a doubly charged
gauge field that should be easily identified if this model is correct. The
other possible channel is $T\rightarrow K^{+}b\rightarrow \nu Eb$.

In the scenario of SUSY 331 models, different channels can be searched. A
good perspective is $pp\rightarrow gg\rightarrow g\rightarrow \widetilde{g}%
\widetilde{g}$ where $\widetilde{g}\rightarrow \widetilde{t}T,t\widetilde{T}%
,T\widetilde{T}$ in the case in which $T$ posseses an ordinary charge $2/3$.
Another production mechanism in the gaugino sector is could be $pp
\rightarrow X^{++}\rightarrow \chi ^{++}\chi ^{0}$, $pp\rightarrow
\chi ^{++}\rightarrow X^{++}\chi ^{0}$, $pp\rightarrow Z,Z^{\prime
++}\chi ^{--}$ \cite{SUSY331 exp}-\cite{SUSY331 exp5}

In looking for unification of the coupling constant by passing through a 331
model, we shall assume that 1) The $331$ gauge group is the only extension
of the SM before the unification of the running coupling constants. 2) The
hypercharge associated with the 331 gauge group is adequately normalized
such that the three gauge couplings unify at certain scale $M_{U}$. and 3)
There is no necessarily a unified gauge group at the scale of convergence of
the couplings $M_{U}$. In the absence of a grand unified group, there are no
restriction on $M_{U}$ coming from proton decay\footnote{%
Notwithstanding, proton decay could be induced even in the absence of a
group of grand unification. It may occurs via six dimensional 331 invariant
effective operators, that violates barionic and leptonic numbers \cite%
{Pleitez effective}.}.

Under our scheme, we have three characteristic energy scales: $M_{U}$ where
the three gauge couplings converge, $M_{X}$ where the $331$ symmetry is
broken, and $M_{Z}$ where the SM breaking occurs. We are going to consider
different scenarios for 331 models with one and three families\footnote{%
331 models with three identical fermion multiplets are usually call one
family models.}, i. e., one family models. We also introduce Supersymmetric
versions of the 331 models with different scalar Higgs multiplets. As for
the SUSY breaking, we shall consider two different scenarios: when SUSY is
broken at the electroweak scale, or when SUSY breaks at the scale of $M_{X}$.

In our scheme we have four parameters to take into account, and to look for
a possible unification of the coupling constants (UCC). They are the scales $%
M_{X}$, $M_{U}$, the value of the coupling contants at the unification
convergence point, $\alpha _{U}$, and the parameter associated with the
normalization of the hypercharge (denoted by $a$). Since we are interested
mostly in possible phenomenological scenarios, the relevant parameter will
be the gauge symmetry breaking scale $M_{X}$; and the parameters $M_{U}$, $a$
can be viewed as functions of this one.

If the unification came from a grand unified symmetry group $G$, the
normalization of the hypercharge $Y$ would be determined by the group
structure. However, under our assumptions, this normalization factor is free
and the problem could be addressed the opposite way, since the values
obtained for $a$ could in turn suggest possible groups of grand unification
in which the 331 group is embedded, we shall explore this possibility as
well.

In the present work, we study six different versions of 331 models with
non-SUSY and SUSY particle content, and find which models could lead to a
unification at certain scale $M_{U}$ with only one symmetry breaking between 
$M_{Z}$ and $M_{U}$ scales. Some of the SUSY versions studied, could provide
a quite natural scenario for split supersymmetry. Finally, we also consider
the possibility of embedding those 331 versions into a grand unified theory
(GUT) in which a gauge group at the unification scale appears.

\section{Running Coupling Constants\label{sec:RCC}}

The evolution for the Running Coupling Constants (RCC) at one-loop order is
ruled by the solution of the Renormalization Group Equations (RGE), which
can be written in the form \cite{quinn}: 
\begin{equation}
\frac{1}{\alpha _{i}(\mu _{2})}=\frac{1}{\alpha _{i}(\mu _{1})}-\frac{b_{i}}{%
2\pi }\ln \left( \frac{\mu _{2}}{\mu _{1}}\right) ,  \label{rge}
\end{equation}%
where $\alpha _{i}=g_{i}^{2}/4\pi $, and the coefficients $b_{i}$ are given
by \cite{Jones}: 
\begin{equation}
b_{i}=\frac{2}{3}\sum_{f}T_{Ri}(f)+\frac{1}{3}\sum_{s}T_{Ri}(s)-\frac{11}{3}%
C_{2i}(G).  \label{betacoe}
\end{equation}%
The summations run over Weyl fermions and scalars, respectively. The
coefficient $T_{R}$ is the Dynkin index 
\begin{equation}
Tr(T^{a}T^{b})_{R}=T_{R}\delta ^{ab},
\end{equation}%
with the generators in the representation R. The last term is the quadratic
Casimir for the adjoint representation 
\begin{equation}
f^{acd}f^{bcd}=C_{2}(G)\delta ^{ab},
\end{equation}%
with $C_{2}(G)=N$ for $SU(N)$. On the other hand, the respective
supersymmetric versions are 
\begin{equation}
b_{i}^{SUSY}=\sum_{f}T_{Ri}(f)+\sum_{s}T_{Ri}(s)-3C_{2i}(G).
\label{betacoesusy}
\end{equation}%
where the usual non-supersymmetric degrees of freedom are counted.

\subsection{Matching Conditions \label{sec:matching}}

The general expression for the electromagnetic charge operator will be a
linear combination of diagonal generators for the gauge group $331$:

\begin{equation}
Q=T_{3}+Y=T_{3}+\frac{2}{\sqrt{3}}bT_{8}+X  \label{Gell-Mann}
\end{equation}%
with $T_{i}$ the Gell-Mann matrices normalized as $Tr(T_{i}T_{j})=\frac{1}{2}%
\delta _{ij}$. The $X$ operator for the abelian group $U(1)_{X}$ is
proportional to the identity matrix $3\times 3$. The hypercharge will be
given by 
\begin{equation}
Y=\frac{2}{\sqrt{3}}bT_{8}+X.  \label{hypercharge1}
\end{equation}%
$b$ is a known parameter that determines the class of 331 models to be
considered \cite{ochoa}.

The renormalization group analysis compares the couplings for different
gauge groups at given energy scales, and models with symmetry breakings need
relations for couplings at different energy regions which are called the
matching conditions; they are extracted from the way in which the unbroken
group is embedded into a bigger broken group. Also, in order to have all
couplings in the same ground, all the generators should be normalized in the
same way, and well normalized couplings are those which will converge in an
unification point.

Calling $\tilde{Y}$ the well-normalized hypercharge operator, it will be
proportional to the original one, 
\begin{equation}
Y=a\tilde{Y},
\end{equation}%
where $a$ is the normalizing parameter so that the convergence of the
running coupling constants at certain scale $M_{U}$ is guaranteed. In the
same way the operator $X$ has a well-normalized $\tilde{X}$, i.e. $X=c\tilde{%
X}$,$\ $and the normalizing parameter for it, is given by Eq. %
\eqref{hypercharge1} requiring the same normalization for $\tilde{Y}$, $%
T_{8} $ which satisfies the following relation 
\begin{equation}
a^{2}=\frac{4}{3}b^{2}+c^{2}.
\end{equation}%
Therefore, the parameter $a$ should be such that 
\begin{equation}
a^{2}\geq \frac{4}{3}b^{2}.  \label{pararestric}
\end{equation}%
Then, the well-normalized hypercharge operator can be written as a function
of the unknown parameter $a$,$\ $as 
\begin{equation}
a\tilde{Y}=\frac{2}{\sqrt{3}}bT_{8}+\sqrt{\left( a^{2}-\frac{4}{3}%
b^{2}\right) }\tilde{X}.  \label{hypercharge}
\end{equation}%
And from this equation, we obtain the following matching condition for the
corresponding couplings \cite{Mohapatra}: 
\begin{equation}
a^{2}\tilde{\alpha}_{Y}^{-1}=\frac{4b^{2}}{3}\alpha _{3L}^{-1}+\left( a^{2}-%
\frac{4}{3}b^{2}\right) \tilde{\alpha}_{X}^{-1}.  \label{matching1}
\end{equation}%
where $\tilde{\alpha}_{Y}$, $\tilde{\alpha}_{X}$, $\alpha _{3L}$ are related
with $U(1)_{\tilde{Y}}$, $U(1)_{\tilde{X}}$ and $SU(3)_{L}$, respectively.

The following relations must also be satisfied 
\begin{eqnarray}
\tilde{\alpha}_{Y} &=&a^{2}\alpha _{Y},\;\;\;\;\tilde{\alpha}_{X}=(a^{2}-%
\frac{4b^{2}}{3})\alpha _{X},  \notag \\
\alpha _{s} &=&\alpha _{3C},\;\;\;\;\alpha _{2L}=\alpha _{3L}.
\label{matching2}
\end{eqnarray}%
where $\alpha _{X}$,$\ \alpha _{Y}$ and $\alpha _{2L}$ are related with $%
U\left( 1\right) _{X},\ U(1)_{Y}$ and $SU(2)_{L}$, respectively. The third
relation corresponds to the strong interaction where $\alpha _{s}$ is
associated with the Standard Model and $\alpha _{3C}$ is related with the
color part in the $331$ model. Finally, the last relation corresponds to the
embedding of $SU(2)_{L}$ into $SU(3)_{L}$.

\subsection{RGE analysis\label{sec:RGE}}

By replacing the relations described by Eqs. (\ref{matching1}, \ref%
{matching2}) into Eq. (\ref{rge}), we can write the evolution for the RCC
from the $Z$ boson-pole $M_{Z}$ passing through a $331$ symmetry breaking
scale $M_{X}$, up to a certain Scale of unification $M_{U}$, 
\begin{eqnarray}
\alpha _{U}^{-1} &=&\frac{1}{a^{2}-\frac{4b^{2}}{3}}\left\{ \alpha
_{EM}(M_{Z})^{-1}-\frac{4b^{2}}{3}\alpha _{2L}(M_{Z})^{-1}\right.  \notag \\
&&\left. \phantom{\frac{1}{a^2-\frac{4b^2}{3}}\frac{1}{\frac{4b^2}{3}}}-%
\frac{b_{Y}-\frac{4b^{2}}{3}b_{2L}}{2\pi }\ln \left( \frac{M_{X}}{M_{Z}}%
\right) -\frac{b_{X}}{2\pi }\ln \left( \frac{M_{U}}{M_{X}}\right) \right\} ,
\label{hiprgene} \\
\alpha _{U}^{-1} &=&\alpha _{2L}(M_{Z})^{-1}-\frac{b_{2L}}{2\pi }\ln \left( 
\frac{M_{X}}{M_{Z}}\right) -\frac{b_{3L}}{2\pi }\ln \left( \frac{M_{U}}{M_{X}%
}\right) ,  \label{debil} \\
\alpha _{U}^{-1} &=&\alpha _{s}(M_{Z})^{-1}-\frac{b_{s}}{2\pi }\ln \left( 
\frac{M_{X}}{M_{Z}}\right) -\frac{b_{3C}}{2\pi }\ln \left( \frac{M_{U}}{M_{X}%
}\right) .  \label{fuerte}
\end{eqnarray}%
where the coefficients $b_{s}$, $b_{2L}$ and $b_{Y}$ are related with $%
SU(3)_{C},SU(2)_{L},U(1)_{Y}$, respectively and they are calculated at
energies in the range $M_{Z}\leq \mu \leq M_{X}$. The coefficients $b_{3C}$, 
$b_{3L},$ and $b_{X}$, are related with $SU(3)_{C},SU(3)_{L},$ and $U(1)_{X}$
respectively; they are calculated for energies in the range $M_{X}\leq \mu
\leq M_{U}$. For our study, we need $b_{i}$ coefficients for energy scales
below (and above) the symmetry breaking scale $M_{X}$, which will be given
by SM (and 331) degrees of freedom. Then for different models, we have
different $b_{i}$'s for the intervals of energy scales which can change the
running of the coupling constants. The input parameters from precision
measurements are \cite{data}%
\begin{eqnarray}
\alpha _{EM}^{-1}(M_{Z}) &=&127.934\pm 0.027,  \notag \\
\sin ^{2}\theta _{w}(M_{Z}) &=&0.23113\pm 0.00015,  \notag \\
\alpha _{s}(M_{Z}) &=&0.1172\pm 0.0020,  \notag \\
\alpha _{2L}^{-1}(M_{Z}) &=&29.56938\pm 0.00068.  \label{input}
\end{eqnarray}

The $M_{U}$ scale, where all the well-normalized couplings have the same
value, can be calculated from \eqref{debil} and \eqref{fuerte} as a function
of the symmetry breaking scale $M_{X}$ 
\begin{equation}
M_{U}=M_{X}\left( \frac{M_{X}}{M_{Z}}\right) ^{-\frac{b_{s}-b_{2L}}{%
b_{3C}-b_{3L}}}\exp \left\{ 2\pi \frac{\alpha _{s}(M_{Z})^{-1}-\alpha
_{2L}(M_{Z})^{-1}}{b_{3C}-b_{3L}}\right\} .  \label{mgut}
\end{equation}%
The hierarchy condition $M_{X}\leq M_{U}\leq M_{Planck}$,$\ $must be
satisfied. We shall however impose a stronger condition of $M_{U}\lesssim
10^{17}$GeV, in order to avoid gravitational effects. Hence, the hierarchy
condition becomes%
\begin{equation}
M_{X}\leq M_{U}\leq 10^{17}GeV  \label{gerarquia}
\end{equation}%
Such condition can establish an allowed range for the symmetry breaking
scale $M_{X}$ in order to obtain grand unification for a given normalizing
parameter $a$.

With a similar procedure, the expression for $a^{2}$ is found, and is given
by 
\begin{eqnarray}
a^{2} &=&\frac{4b^{2}}{3}+\left\{ \alpha _{EM}(M_{Z})^{-1}-\frac{4b^{2}}{3}%
\alpha _{2L}(M_{Z})^{-1}-\frac{b_{Y}-\frac{4b^{2}}{3}b_{2L}}{2\pi }\ln
\left( \frac{M_{X}}{M_{Z}}\right) \right.  \notag \\
&+&\left. b_{X}\left[ \frac{1}{2\pi }\frac{b_{s}-b_{2L}}{b_{3C}-b_{3L}}\ln
\left( \frac{M_{X}}{M_{Z}}\right) -\frac{\alpha _{s}(M_{Z})^{-1}-\alpha
_{2L}(M_{Z})^{-1}}{b_{3C}-b_{3L}}\right] \right\}  \notag \\
&\times &\left\{ \alpha _{2L}(M_{Z})^{-1}-\frac{b_{2L}}{2\pi }\ln \left( 
\frac{M_{X}}{M_{Z}}\right) +b_{3L}\left[ \frac{1}{2\pi }\frac{b_{s}-b_{2L}}{%
b_{3C}-b_{3L}}\ln \left( \frac{M_{X}}{M_{Z}}\right) \right. \right.  \notag
\\
&-&\left. \left. \frac{\alpha _{s}(M_{Z})^{-1}-\alpha _{2L}(M_{Z})^{-1}}{%
b_{3C}-b_{3L}}\right] \right\} ^{-1}  \notag \\
&\geq &\frac{4}{3}b^{2}.  \label{ados}
\end{eqnarray}

In order to analyze the possibility of having unification at certain scale $%
M_{U}\ $with $331\ $as the only intermediate gauge group between the SM and $%
M_{U}$ scales, we could distinguish three scenarios of unification pattern

\begin{dingautolist}{202}
\item The scenario with $b_{3C}\neq b_{3L}$ and $\left( b_{3C}-b_{3L}\right)
\neq \left( b_{s}-b_{2L}\right) $. We shall call it the first unification
pattern (\textbf{1UP}). In that case, we obtain an allowed region for the
scale$\ M_{X}$. It is carried out by combining the results of Section \ref%
{sec:RGE}. The procedure to get the allowed interval for $M_{X}$ is
described in detail in Section \ref{sec:A} for the so called model A.

\item The scenario with $b_{3C}=b_{3L}$. We call it (\textbf{2UP}). In such
a case Eq. (\ref{mgut}) is not valid anymore, and we should go back to Eqs. (%
\ref{debil}, \ref{fuerte}). For an arbitrary value of $M_{X}$ the couplings $%
\alpha _{3C}$ and $\alpha _{3L}$ go parallel each other for energies larger than $M_{X}$.
Therefore, the only possible way to still obtain unification is by setting $\alpha _{3C}=\alpha _{3L}$ at the scale $M_{X}$
such that both couplings go together for scales above $M_{X}$. Unification with the third coupling could occur at
any scale bigger than $M_{X}$.  By equating Eqs
(\ref{debil}, \ref{fuerte}), we find a single value of $M_{X}$ that makes
the couplings$\ \alpha _{3C}$ and $\alpha _{3L}$ to converge. This
convergence occurs at the scale%
\begin{equation}
M_{X}=M_{Z}\exp \left[ \frac{2\pi \left[ \alpha _{2L}(M_{Z})^{-1}-\alpha
_{s}(M_{Z})^{-1}\right] }{\left( b_{2L}-b_{s}\right) }\right]  \label{Mx}
\end{equation}%
It worths emphasizing that this scenario leads to a unique value of $M_{X}$ and not
to an allowed range. Finally, Eq. (\ref{ados}) for $a^{2}$ must also be
recalculated to find%
\begin{eqnarray}
a^{2} &=&\frac{F_{1}\left( M_{X}\right) -\frac{b_{X}}{2\pi }\ln \left( \frac{%
M_{U}}{M_{X}}\right) }{F_{2}\left( M_{X}\right) -\frac{b_{3}}{2\pi }\ln
\left( \frac{M_{U}}{M_{X}}\right) }+\frac{4b^{2}}{3}  \notag \\
F_{1}\left( M_{X}\right) &=&\alpha _{EM}(M_{Z})^{-1}-\frac{4b^{2}}{3}\alpha
_{2L}(M_{Z})^{-1}-\frac{b_{Y}-\frac{4b^{2}}{3}b_{2L}}{2\pi }\left\{ \frac{%
2\pi \left[ \alpha _{2L}(M_{Z})^{-1}-\alpha _{s}(M_{Z})^{-1}\right] }{\left(
b_{2L}-b_{s}\right) }\right\}  \notag \\
F_{2}\left( M_{X}\right) &=&\alpha _{2L}(M_{Z})^{-1}-\frac{b_{2L}}{2\pi }%
\left\{ \frac{2\pi \left[ \alpha _{2L}(M_{Z})^{-1}-\alpha _{s}(M_{Z})^{-1}%
\right] }{\left( b_{2L}-b_{s}\right) }\right\}  \label{a2N}
\end{eqnarray}

\item The \textbf{3UP} with $\left( b_{3C}-b_{3L}\right) =\left(
b_{s}-b_{2L}\right) \neq 0$;$\ $according to Eq. (\ref{mgut}), the
unification scale $M_{U}$ becomes independent of $M_{X}$.
\end{dingautolist}

The case $\left( b_{3C}-b_{3L}\right) =\left( b_{s}-b_{2L}\right) =0$, does
not lead to unification as can be seen by trying to equate Eqs. (\ref{debil}%
) and (\ref{fuerte}). Since the first scenario is the most commom one, we
shall only indicate when the other two scenarios appear. We will study
non-SUSY and SUSY versions of the 331 models. In the case of SUSY models we
shall consider two scenarios for the SUSY breaking pattern

\begin{enumerate}
\item The SUSY Breaking Scenario at the $Z-$pole (\textbf{ZSBS}), in which
the SUSY breaking scale is taken as $\Lambda _{SUSY}\sim M_{Z}$. Although
this is not a very realistic scenario, numerical results do not change
significantly with respect to the more realistic scenario with $\Lambda
_{SUSY}$ lying at some few TeV's.

\item The SUSY Breaking Scenario at the $M_{X}$ scale (\textbf{XSBS}), with $%
\Lambda _{SUSY}\sim M_{X}$ i.e. SUSY breaking at the 331 breaking scale.
\end{enumerate}

\section{$331$ Models}

Analogously to the SM, fermions will transform as singlets or in the
fundamental representation of $SU(3)_{L}$, and gauge fields in the adjoint
representation. Assignment of $U(1)_{X}$ quantum numbers should be done
ensuring a model free of anomalies. There are models with three different
families necessary to cancel out the anomalies, and there are models with
only one family and the other two are a copy of the first. We will take into
account six different versions of the 331 model for the analysis of the
unification scheme.


The minimal spectrum necessary for symmetry breaking and generation of
masses is given by \cite{ochoa1} 
\begin{equation}
\begin{aligned} \phi_{1}&\sim (1,3^*,-1/3),\\ \phi_{2}&\sim (1,3^*,-1/3),\\
\phi_{3}&\sim (1,3^*,2/3). \end{aligned}
\end{equation}%
Where the quantum numbers are associated with $SU\left( 3\right) _{C},\
SU\left( 3\right) _{L},\ $and $U\left( 1\right) _{X}$ respectively. The
first multiplet acquires a vacuum expectation value (VEV) at $M_{X}$ scale,
breaking the symmetry as $331\rightarrow 321$; the other two will be
decomposed as singlets and doublets of $SU(2)_{L}$, which will break the $%
321 $ symmetry. The spectrum transforms like 
\begin{eqnarray}
\phi _{2} &\rightarrow &%
\begin{cases}
\phi _{2SM} & \sim (1,2^{\ast },-1/2), \\ 
\phi _{2}^{0} & \sim (1,1,0).%
\end{cases}
\\
\phi _{3} &\rightarrow &%
\begin{cases}
\phi _{3SM} & \sim (1,2^{\ast },1/2), \\ 
\phi _{2}^{+} & \sim (1,1,1).%
\end{cases}%
\end{eqnarray}%
This is applied for models with $b=1/2$, with tiny changes that do not
affect our analysis \cite{Ponce1,Ponce2}. For models with $b=3/2$ this
spectrum should be varied in order to get a phenomenological mass spectrum.
For this case we have the following transformation for the triplets \cite%
{Pleitez, Frampton, Pleitez2, Pleitez3}: 
\begin{equation}
\begin{aligned} \eta&\sim (1,3^*,0),\\ \rho&\sim (1,3^*,1),\\ \chi&\sim
(1,3^*,-1). \end{aligned}
\end{equation}%
the $331$ symmetry is broken when $\chi $ acquires a VEV in the third
component. The remaining scalars will transform in the $321$ symmetry as 
\begin{eqnarray}
\eta &\rightarrow &%
\begin{cases}
\eta _{SM} & \sim (1,2^{\ast },-1/2), \\ 
\eta _{2}^{+} & \sim (1,1,1).%
\end{cases}
\\
\rho &\rightarrow &%
\begin{cases}
\rho _{SM} & \sim (1,2^{\ast },1/2), \\ 
\rho ^{++} & \sim (1,1,2).%
\end{cases}%
\end{eqnarray}

The Model that we will call Model $E$ needs also another scalar boson
transforming under $\underline{6}$ representation of $SU(3)_{L}$. Such an
structure is required in order to give masses to the neutrinos \cite%
{Pleitez, Frampton, Pleitez2}, 
\begin{equation}
S\sim (1,6,0)
\end{equation}%
which has the following representation in the $321$ symmetry: 
\begin{equation}
S\rightarrow 
\begin{cases}
\phi _{3SM} & \sim (1,3,1), \\ 
\phi _{2SM} & \sim (1,2,1/2), \\ 
\phi ^{--} & \sim (1,1,-2).%
\end{cases}%
\end{equation}

\subsection{Model A\label{sec:A}}

\begin{figure}[tbp]
\begin{center}
\psfrag{B}{\hspace{-15mm}$M_{U}(GeV)$} 
\psfrag{A}{\vspace{5mm}{\small
$M_{X}(GeV)$}} \includegraphics[scale=0.6]{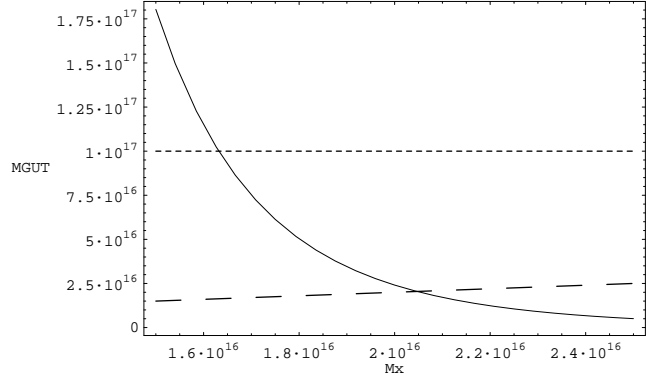}
\end{center}
\caption{Allowed interval for $M_{X}$ in model A. The solid line represents $%
M_{U}$ as a function of $M_{X}$. The horizontal short dashed line is the
graphics for $M_{U}=10^{17}$GeV. The long dash line represents the function $%
M_{U}=M_{X}$.}
\label{fig:Mx}
\end{figure}

\begin{figure}[tbp]
\begin{center}
\psfrag{B}{\hspace{-4mm}$a^{2}$} 
\psfrag{A}{\vspace{5mm}{\small
$M_{X}(GeV)$}} \includegraphics[scale=0.6]{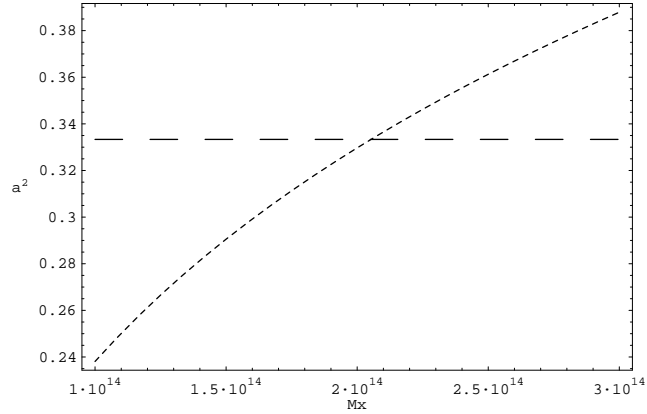}
\end{center}
\caption{Normalization parameter $a^{2}$ as function of the symmetry
breaking scale $M_{X}$ for model A. The intersection of this function with
the constant function $\frac{4}{3}b^{2}$ gives a lower limit for $M_{X}$
owing to the condition described by Eq. (\protect\ref{pararestric}).}
\label{fig:a2}
\end{figure}
\begin{figure}[tbph]
\begin{center}
\psfrag{B}{\hspace{-7mm}$\alpha_{U}^{-1}$} \psfrag{A}{\hspace{-3mm}%
\vspace{-40mm}$\mu(GeV)$} \includegraphics[scale=0.6]{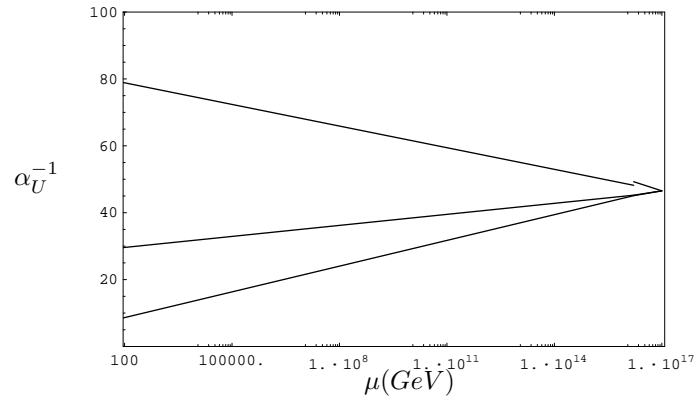}
\end{center}
\caption{Running Coupling Constants for model A with $M_{X}=1.63\times
10^{16}GeV$. The unification scale appears at $M_{U}\simeq 10^{17}$GeV}
\label{fig:RCC}
\end{figure}

The simplest anomaly-free structure for this gauge group is generated by
taking $b=1/2$; and the following spectrum is obtained \cite{Martinez}\label%
{Martinez} 
\begin{eqnarray}
\psi _{1L} &=&(e^{-},\nu _{e},N_{1}^{0})_{L}^{T}\sim (1,3^{\ast },-1/3), 
\notag  \label{331eq} \\
\psi _{2L} &=&(E^{-},N_{2}^{0},N_{3}^{0})_{L}^{T}\sim (1,3^{\ast },-1/3), 
\notag \\
\psi _{3L} &=&(N_{4}^{0},E^{+},e^{+})_{L}^{T}\sim (1,3^{\ast },2/3),  \notag
\\
Q_{L} &=&(u,d,D)_{L}^{T}\sim (3,3,0), \\
u_{L}^{c} &\sim &(3^{\ast },1,-2/3),\;\;\;\;d_{L}^{c}\sim (3^{\ast
},1,1/3),\;\;\;\;D_{L}^{c}\sim (3^{\ast },1,1/3).  \notag
\end{eqnarray}%
The other two families are copies of the first one, and each family is free
of anomalies. This particular 331 model could be embbeded in the $E_{6}$
gauge theory, but we shall assume that not necessarily such embedding occurs.

From the general expression for coefficients $b_{i}$ Eq. \eqref{betacoe},
and using the quantum numbers assigned for each representation in this
model, they can be expressed at energies below $M_{X}$ in the following form 
\begin{eqnarray}
b_{Y} &=&\frac{20}{9}N_{g}+\frac{1}{6}N_{H}+\frac{1}{3}\sum_{sing-s}Y^{2}(s),
\notag \\
b_{2L} &=&\frac{4}{3}N_{g}+\frac{1}{6}N_{H}-\frac{22}{3},  \notag \\
b_{s} &=&\frac{4}{3}N_{g}-11.
\end{eqnarray}%
where $N_{g}$ is the number of families, $N_{H}$ the number of scalar
doublets in $SU(2)_{L}$, and the sum runs over scalar singlets.

In the same way, the $b_{i}$'s for energies above the $M_{X}$ are given by 
\begin{eqnarray}
b_{X} &=&\frac{8}{3}N_{g}+\frac{2}{3},  \notag \\
b_{3L} &=&2N_{g}+\frac{1}{2}-11,  \notag \\
b_{3C} &=&2N_{g}-11.
\end{eqnarray}%
Therefore, from the spectrum showed in Eq. \eqref{331eq}, we obtain 
\begin{equation}
(b_{Y},b_{2L},b_{s})=\left( \frac{22}{3},-3,-7\right)  \label{betaYA}
\end{equation}%
for energies below $M_{X}$, and 
\begin{equation}
(b_{X},b_{3L},b_{3C})=\left( \frac{26}{3},-\frac{9}{2},-5\right)
\label{betaXA}
\end{equation}%
for energies above $M_{X}$.

First of all, by using the hierarchy condition Eq. (\ref{gerarquia}), an
allowed region for $M_{X}$ can be found. In Fig. \ref{fig:Mx} we plot $M_{U}$
as a function of $M_{X}$, Eq. (\ref{mgut}). We also plot the line $%
M_{U}=M_{X}$,\ and\ the\ line\ $M_{U}=10^{17}$GeV, the intersection of these
lines with the first plot, provides the allowed region given by the
constraints $M_{X}\leq M_{U}\leq 10^{17}$GeV according to Eq. (\ref%
{gerarquia}). The intersection between the curve $M_{U}(M_{X})$ from Eq. (%
\ref{mgut}) with $M_{U}=10^{17}$ GeV gives the lower bound for $M_{X}$, and
the intersection with $M_{U}=M_{X}$ gives the upper limit for $M_{X}$. Then
from Fig. \ref{fig:Mx} and the hierarchy condition, it is obtained that 
\begin{equation}
1.63\times 10^{16}{\text{G}eV}\leq M_{X}\leq 2.05\times 10^{16}{\text{G}eV}.
\label{intervalXA}
\end{equation}%
In Figure \ref{fig:a2} we plot $a^{2}$ as a function of $M_{X}$ from Eq. (%
\ref{ados}). For this model $b=1/2$ and from Eq. (\ref{pararestric}) the
restriction $a^{2}\geq 1/3$ is obtained. The horizontal line $a^{2}=1/3$ is
also showed and the allowed values are above of this line. In this case the
constraint for $M_{X}$ is given by $M_{X}\leq 2\times 10^{14}$ GeV which is
weaker than the previous restriction.

From Eq. (\ref{mgut}) and the allowed interval for $M_{X}$ of Eq. (\ref%
{intervalXA}) we get the region permitted for $M_{U}$ i.e. $2\times 10^{16}$
GeV$\ \leq M_{U}\ \leq \ 10^{17}$GeV.

Now, for $M_{X}=2.05\times 10^{16}$GeV (the highest allowed value) we obtain
from Eq. (\ref{mgut}) that $M_{U}\approx M_{X}$ which is not a natural
hierarchy. Instead, for $M_{X}=1.63\times 10^{16}{\text{G}}${eV (the lowest
allowed value)} the unification occurs at a more natural scale of $10^{17}$%
GeV. Then, we shall take the latter to plot the evolution of the running
coupling constants.

The evolution of the RCC described by Eqs. (\ref{rge}), requires the results
for the matching conditions Eqs. (\ref{matching1}, \ref{matching2}).
Moreover, the following input parameters are necessary: \ding{172} The $%
b_{i} $ coefficients Eqs. (\ref{betaYA}, \ref{betaXA}), \ding{173} The value
of the 331 breaking scale $M_{X}$, and \ding{174} the values of the
couplings at Z-pole Eqs. (\ref{input}). Assuming the lowest allowed value
for the 331 breaking scale ($M_{X}=1.63\times 10^{16}{\text{G}}${eV}) we
plot the evolution of the running coupling constants in Fig. \ref{fig:RCC}.
The unification scale appears to be $M_{U}=10^{17}$GeV, in agreement with
our previous results.

\subsubsection{SUSY Extension}

One possible SUSY extension in which the superpartners cancel the anomalies
of their partners, arises by introducing new scalars which transform in the
conjugate representations of the original ones. Under these considerations,
and using Eqs. (\ref{betacoesusy}) we obtain the supersymmetric $b_{i}$
coefficients 
\begin{eqnarray}
(b_{Y},b_{2L},b_{s}) &=&\left( 14,2,-3\right) ,  \label{betaYASUSYII} \\
(b_{X},b_{3L},b_{3C}) &=&\left( 16,3,0\right) .  \label{betaXASUSYII}
\end{eqnarray}%
We take first the \textbf{ZSBS} in which SUSY breaking is assumed to occur
at EW scale \footnote{%
Our results are also valid for SUSY breaking at scales of the order of $1$
TeV.}. Since in that case SUSY is not broken by going from EW to the
unification scale, we should use the supersymmetric $b_{i}$ coefficients, at
scales below and above the 331 breaking scale. Taking it into account, and
using the hierarchy condition (\ref{gerarquia}), as it was explained in the
previous section, we get an allowed range for $M_{X}$, of%
\begin{equation}
1.27\times 10^{8}{\text{G}}\text{{eV}}\leq M_{X}\leq 2.76\times 10^{13}{%
\text{G}}\text{{eV}}.  \label{interval SUSY A1}
\end{equation}%
The bound coming from the $a$ parameter Eq. (\ref{pararestric}), yields $%
6.43\times 10^{6}{\text{G}eV}\leq M_{X}$, \ which gives no further
restriction. This allowed region for $M_{X}$ leads to an allowed interval
for $M_{U}$ obtained from Eq. (\ref{mgut}), getting\footnote{%
When a group of grand unification is present, protection from proton decay
leads to a stronger bound for $M_{X}$ by demanding $M_{U}\gtrsim 2\times
10^{16}$GeV, in that case we find $1.27\times 10^{8}{\text{G}}${eV}$\leq
M_{X}\leq 1.4\times 10^{9}{\text{G}}${eV}.}%
\begin{equation*}
2.76\times 10^{13}\text{GeV}\leq M_{U}\leq 10^{17}\text{GeV}
\end{equation*}

On the other hand, we shall consider the \textbf{XSBS} with SUSY breaking at
the $M_{X}$ scale. In that case, we should use the non-SUSY $b_{i}$
coefficients for scales below $M_{X}$, and the SUSY $b_{i}$ coefficients for
energies above $M_{X}$. From this we obtain the following allowed interval
for $M_{X}$ 
\begin{equation}
1.76\times 10^{14}{\text{G}}\text{{eV}}\leq M_{X}\leq 2.05\times 10^{16}{%
\text{G}}\text{{eV}},
\end{equation}%
for which the restriction $a^{2}>1/3$ is also satisfied. This allowed region
provides a unification scale between $2.05\times 10^{16}GeV$ and $10^{17}GeV$
which correspond to the values calculated from Eq. (\ref{mgut}) for the
lower and upper values for $M_{X}$, respectively. Therefore, this is a
possible scenario of unification with 331 as the unique intermediate gauge
group, for both schemes of SUSY breaking.

It worths noting that SUSY scenarios leads to much lower breaking scales for
the 331 model, especially for low energy SUSY breaking.

In another possible supersymmetric extension of model A, the scalar fields
are not introduced explicitly. Instead, they appear as superpartners of the
leptons, since they have the same representations and quantum numbers \cite%
{Martinez3}. The supersymmetric $b_{i}$ coefficients read%
\begin{eqnarray}
(b_{Y},b_{2L},b_{s}) &=&\left( 10,0,-3\right) ,  \label{betaYASUSYI} \\
(b_{X},b_{3L},b_{3C}) &=&\left( 12,0,0\right) .  \label{betaXASUSYI}
\end{eqnarray}%
For both schemes of SUSY breaking, the model correspond to the \textbf{2UP}
explained in Sec. (\ref{sec:RGE}), so that $M_{X}$ acquires a single value.
The 331 scale appears at $M_{X}\simeq 10^{21}$GeV. It discards this SUSY
version of model A, to get \textbf{UCC} under our scheme.

\subsection{Model B}

Again the parameter $b$ takes the value $1/2$, and in this case the 331
gauge theory is a family-symmetric model. For each family the spectrum is 
\cite{Martinez2}: 
\begin{eqnarray}
\psi _{1L} &=&(e^{-},\nu _{e},E_{1}^{-})_{L}^{T}\sim (1,3^{\ast },-2/3), 
\notag \\
\psi _{2L} &=&(N1^{0},E_{2}^{+},\nu _{e}^{c})_{L}^{T}\sim (1,3^{\ast },1/3),
\notag \\
\psi _{3L} &=&(E_{2}^{-},N_{2}^{0},E_{3}^{-})_{L}^{T}\sim (1,3^{\ast },2/3),
\notag \\
e_{L}^{+} &\sim &(1,1,1),\;\;\;\;E_{1L}^{+}\sim
(1,1,1),\;\;\;\;E_{3L}^{+}\sim (1,1,1),  \notag \\
Q_{L} &=&(u,d,U)_{L}^{T}\sim (3,3,1/3), \\
u_{L}^{c} &\sim &(3^{\ast },1,-2/3),\;\;\;\;d_{L}^{c}\sim (3^{\ast
},1,1/3),\;\;\;\;U_{L}^{c}\sim (3^{\ast },1,-2/3).  \notag
\end{eqnarray}%
In this case, the 331 gauge group could be embbeded into the $SU(6)\otimes
U(1)_{X}$ gauge theory, but we suppose it is not mandatory. This model has
the same spectrum of model A at low energies. Therefore, it has the same
values for the $b_{i}$ coefficients at low energies given by Eq. (\ref%
{betaYA}) 
\begin{equation}
(b_{Y},b_{2L},b_{s})=\left( \frac{22}{3},-3,-7\right) .
\end{equation}%
For energies larger than $M_{X}$, we have that $b_{3L}$ and $b_{3C}$ are the
same as in model A, since their values are ruled by the number of triplets,
which coincide for both models. For $b_{X}$, there is a difference coming
from the additional singlets in the lepton sector. In this case, we obtain 
\begin{equation}
b_{X}=\frac{20}{3}N_{g}+\frac{2}{3}=\frac{62}{3}.  \notag
\end{equation}%
Hence, 
\begin{equation}
(b_{X},b_{3L},b_{3C})=\left( \frac{62}{3},-\frac{9}{2},-5\right) .
\label{betaXB}
\end{equation}

Following the same procedure as for the Model A, we find the allowed range
for $M_{X}$%
\begin{equation}
1.63\times 10^{16}{\text{G}eV}\leq M_{X}\leq 2.05\times 10^{16}{\text{G}eV}.
\end{equation}%
In this interval,$\ a^{2}>1/3$ yields $M_{X}\geq 3.42\times 10^{15}$ GeV,
and $M_{U}$ lies in the interval $\left[ 2.05\times 10^{16},10^{17}\right] $%
GeV.

\subsubsection{SUSY Extension}

As a first attempt, we take the model in Ref. \cite{Martinez4}, where
scalars are considered as sleptons and two new superfields are introduced in
order to give mass to the supersymmetric particles and cancel out the
anomalies. They transform as 
\begin{equation}
\hat{\psi}_{4L}\sim (1,3,1/3),\quad \hat{\psi}_{5L}\sim (1,3^{\ast },-1/3).
\end{equation}%
The spectrum is the same at low energies as SUSY model A, Eq. (\ref%
{betaYASUSYI}). The supersymmetric $b_{i}$ coefficients read 
\begin{subequations}
\begin{align}
(b_{Y},b_{2L},b_{s})& =\left( 10,0,-3\right) ,  \label{betaYBSUSYI} \\
(b_{X},b_{3L},b_{3C})& =\left( 24,3,0\right) .  \label{betaXBSUSYI}
\end{align}%
For the \textbf{ZSBS} ($\Lambda _{SUSY}\approx M_{Z}$), we are in the 
\textbf{3UP} explained in Sec. (\ref{sec:RGE}) so that $M_{U}$ acquires a
single value 
\end{subequations}
\begin{equation}
M_{U}=1.24\times 10^{21}GeV.
\end{equation}%
which discards this SUSY version of model B with $\Lambda _{SUSY}\approx
M_{Z}$, to achieve unification with our breaking scheme.

In the \textbf{XSBS} ( $\Lambda _{SUSY}\approx M_{X}$), the allowed region
for the model is 
\begin{equation}
1.76\times 10^{14}{\text{G}eV}\leq M_{X}\leq 2.05\times 10^{16}\text{GeV}.
\end{equation}%
the bound from $a^{2}$ is $M_{X}\geq 1.44\times 10^{11}$GeV giving no
further restriction. $M_{U}\ $belongs to the interval $\left[ 2.05\times
10^{16},\ 10^{17}\right] $GeV. Hence, this SUSY version of model B with $%
\Lambda _{SUSY}\approx M_{X}$ permits \textbf{UCC} under our assumptions.

\subsection{Model C}

In this model, the parameter $b$ is also equal to 1/2, but the cancellation
of anomalies is obtained with a number of families multiple of three \cite%
{Ozer}. Its fermionic spectrum is 
\begin{eqnarray}
\psi _{L}^{\alpha } &=&(\nu _{\alpha },{\alpha }^{-},E_{\alpha
}^{-})_{L}^{T}\sim (1,3,-2/3),  \notag \\
\alpha _{L}^{+} &\sim &(1,1,1),\;\;\;\;E_{\alpha L}^{+}\sim (1,1,1),  \notag
\\
Q_{L}^{i} &=&(d^{i},u^{i},U^{i})_{L}^{T}\sim (3,3^{\ast },1/3),  \notag \\
u_{L}^{ic} &\sim &(3^{\ast },1,-2/3),\;\;\;\;d_{L}^{ic}\sim (3^{\ast
},1,1/3),\;\;\;\;U_{L}^{ic}\sim (3^{\ast },1,-2/3),  \notag \\
Q_{L}^{3} &=&(u_{3},d_{3},D)_{L}^{T}\sim (3,3,0), \\
u_{3L}^{c} &\sim &(3^{\ast },1,-2/3),\;\;\;\;d_{3L}^{c}\sim (3^{\ast
},1,1/3),\;\;\;\;D_{L}^{c}\sim (3^{\ast },1,/3).  \notag
\end{eqnarray}%
where $\alpha =1,2,3$ label the three lepton families; while $i=1,2$ refer
to only two of the quark families because the third one transforms in a
different way. The spectrum is the same of the Model A at low energies, so
that the $b_{i}$ coefficients are given by Eq. (\ref{betaYA}). For energies
above $M_{X}$ there are new contributions and the $b_{i}$'s yield 
\begin{eqnarray}
(b_{Y},b_{2L},b_{s}) &=&\left( \frac{22}{3},-3,-7\right)  \notag \\
(b_{X},b_{3L},b_{3C}) &=&\left( \frac{42}{3},-\frac{13}{2},-5\right) .
\end{eqnarray}%
Following the same procedure explained before, we find that the allowed
region for $M_{X}$ is 
\begin{equation}
2.05\times 10^{16}GeV\leq M_{X}\leq 3.15\times 10^{16}GeV.  \label{3.30}
\end{equation}%
In this case $a^{2}\geq 1/3$, leads to $M_{X}\leq 9.57\times 10^{18}$GeV and
provides an upper (and not a lower) limit because $a^{2}$ is a decreasing
function of $M_{X}$. Of course, this constraint gives no further
restrictions. Taking values for $M_{X}$ from the interval in Eq. (\ref{3.30}%
), the permitted interval for $M_{U}$ is$\ \left[ 2.05\times 10^{16}\text{GeV%
},10^{17}\text{GeV}\right] $.

\subsubsection{SUSY Extension}

The extension is done adding new scalars transforming in the conjugate
representation in order to cancel the anomalies. With this spectrum we
obtain 
\begin{eqnarray}
(b_{Y},b_{2L},b_{s}) &=&\left( 14,2,-3\right)  \notag \\
(b_{X},b_{3L},b_{3C}) &=&\left( 24,0,0\right) .  \label{betaXCSUSY}
\end{eqnarray}%
At low energy scales, the coefficients coincide with the analogous in model
A, Eq. (\ref{betaYASUSYII}). For both schemes of SUSY breaking we lie in\
the \textbf{2UP} getting a unique $M_{X}$ from Eq. (\ref{Mx}). For the 
\textbf{ZSBS}, $M_{X}$ becomes%
\begin{equation}
M_{X}=2.76\times 10^{13}GeV.
\end{equation}%
If we fit $M_{U}=2.76\times 10^{13}\left( =1\times 10^{17}\right) $ GeV, we
need $a^{2}=3.26\left( =1.78\right) $ to obtain unification of the three
couplings according to Eq. (\ref{a2N}). In general, for $M_{U}$ lying in the
interval $\left[ 2.76\times 10^{13}\text{GeV},\ 10^{17}\text{ GeV}\right] $,
we find that the $a^{2}\ $parameter lies in the interval $\left[ 1.78,\ 3.26%
\right] $; and the condition $a^{2}\geq \left( 4/3\right) b^{2}$ is
satisfied. Therefore, this model gives a consistent unification with 331 as
the only intermediate gauge group.

For the \textbf{XSBS}, $M_{X}=2.05\times 10^{16}\ $GeV. From $M_{U}\in \left[
2.05\times 10^{16},\ 1\times 10^{17}\right] $ GeV, we get $a^{2}\in \left[
1.84,\ 1.97\right] $ and $a^{2}\geq \left( 4/3\right) b^{2}$ is fulfilled in
all cases and the model is viable as well.

\subsection{Model D}

This model differs from model C just in its spectrum \cite{Long, Long2},
which is given by 
\begin{eqnarray}
\psi _{L}^{\alpha } &=&({\alpha }^{-},\nu _{\alpha },N_{\alpha
}^{0})_{L}^{T}\sim (1,3,-1/3),  \notag \\
\alpha _{L}^{+} &\sim &(1,1,-1),  \notag \\
Q_{L}^{i} &=&(d^{i},-u^{i},D^{i})_{L}^{T}\sim (3,3^{\ast },0),  \notag \\
u_{L}^{ic} &\sim &(3,1,+2/3),\;\;\;\;d_{L}^{ic}\sim
(3,1,-1/3),\;\;\;\;D_{L}^{ic}\sim (3,1,-1/3),  \notag \\
Q_{L}^{3} &=&(d_{3},u_{3},U)_{L}^{T}\sim (3,3,1/3), \\
u_{3L}^{c} &\sim &(3,1,2/3),\;\;\;\;d_{3L}^{c}\sim
(3,1,-1/3),\;\;\;\;U_{L}^{c}\sim (3,1,2/3).  \notag
\end{eqnarray}%
with $\alpha =1,2,3$, running over the three families while $i=1,2$ runs
over two families. The model differs from the others above in its spectrum
at high energies. Therefore for the $331$ gauge group the $b_{i}$ functions
coincide with the values in the Eq. (\ref{betaYA}), 
\begin{equation}
(b_{Y},b_{2L},b_{s})=\left( \frac{22}{3},-3,-7\right) .
\end{equation}%
For energies larger than the symmetry breaking scale $M_{X}$ we have 
\begin{equation}
(b_{X},b_{3L},b_{3C})=\left( \frac{26}{3},-\frac{13}{2},-5\right) .
\label{betaXD}
\end{equation}%
These values give the following allowed region 
\begin{equation}
2.05\times 10^{16}GeV\leq M_{X}\leq 3.15\times 10^{16}GeV,
\end{equation}%
and $a^{2}>1/3$ predicts $M_{X}\leq 1.54\times 10^{20}$GeV, yielding no
further restriction. The scale of grand unification lies in the interval $%
2.05\times 10^{16}$GeV$\leq M_{U}\leq 10^{17}$GeV.

\subsubsection{SUSY Extension}

Treating the model in the same way as model C, the $b_{i}$ coefficients at
low energies are given in Eq. (\ref{betaYASUSYII}), and for high energy
scales they are given by 
\begin{equation}
(b_{X},b_{3L},b_{3C})=\left( 16,0,0\right) .  \label{betaXDSUSY}
\end{equation}%
Both schemes of SUSY breaking are in the \textbf{2UP}. For the \textbf{ZSBS}
we find $M_{X}=2.76\times 10^{13}GeV$. Taking$\ M_{U}\in \left[ 2.76\times
10^{13},\ 1\times 10^{17}\right] $ GeV we obtain $a^{2}\in \left[ 2.27,\ 3.26%
\right] $, and the basic restriction on $a^{2}$ is always accomplished. With 
\textbf{XSBS}, it is found that $M_{X}=2.05\times 10^{16}$, and we get $%
M_{U}\in \left[ 2\times 10^{16},\ 1\times 10^{17}\right] $ GeV for $a^{2}=%
\left[ 1.88,\ 1.97\right] $, respectively.

Thus the SUSY version provides a possible scenario of unification for both
schemes of SUSY breaking.

\subsection{Model E}

In this model $b=3/2$ and the cancellation of anomalies is obtained by the
interplay of three families \cite{Pleitez, Frampton, Pleitez2}. Its
fermionic spectrum is 
\begin{eqnarray}
\psi _{L}^{\alpha } &=&(e^{\alpha },\nu ^{\alpha },e^{c{\alpha }%
})_{l}^{T}\sim (1,3^{\ast },0),  \notag \\
q_{L}^{i} &=&(u^{i},d^{i},j^{i})_{L}^{T}\sim (3,3,-1/3),  \notag \\
q_{L}^{1} &=&(d^{1},u^{1},s)_{L}^{T}\sim (3,3^{\ast },2/3),  \notag \\
u_{L}^{c{\alpha }} &\sim &(3,1,-2/3),\;\;\;\;d_{L}^{c{\alpha }}\sim
(3,1,1/3),\;\;\;\;s_{L}^{c}\sim (3,1,-5/3),  \notag \\
j_{L}^{ci} &\sim &(3,1,4/3).
\end{eqnarray}%
where ${\alpha }=1,2,3$ labels the families, and $i=2,3$ are related with
two of them.

Taking into account the whole possible scalar spectrum at low energy scales,
including the scalars in the 6-dimensional representation, we can write the $%
b_{i}$ coefficients as 
\begin{eqnarray}
b_{Y} &=&\frac{20}{9}N_{g}+\frac{1}{6}N_{H}+\frac{1}{3}3Y^{2}(triplet)+\frac{%
1}{3}\sum_{singl-s}Y^{2}(s),  \notag  \label{MEGENE} \\
b_{2L} &=&\frac{4}{3}N_{g}+\frac{1}{6}N_{H}+\frac{1}{3}T_{R}(triplet)-\frac{%
22}{3},  \notag \\
b_{s} &=&\frac{4}{3}N_{g}-11.
\end{eqnarray}%
where $Y(triplet)$ and $T_{R}(triplet)$ mean the quantum numbers of the $%
SU(2)_{L}$ scalar triplet in the 6-dimensional representation. Therefore the
values of the $b_{i}$ coefficients read%
\begin{eqnarray}
(b_{Y},b_{2L},b_{s}) &=&\left( \frac{23}{2},-\frac{13}{6},-7\right) ,
\label{betaYE} \\
(b_{X},b_{3L},b_{3C}) &=&\left( 22,-\frac{17}{3},-5\right) .  \label{betaXE}
\end{eqnarray}

With these $b_{i}$ coefficients we only find a lower limit from the
hierarchy condition, 
\begin{equation}
6.87\times 10^{13}GeV\leq M_{X}
\end{equation}%
An upper limit is obtained from Eq. (\ref{pararestric}) with $3\leq a^{2}$,
which is satisfied for $M_{X}\leq 6.03\times 10^{12}GeV$, but both
constraints are inconsistent and rule out the model under our assumptions.

\subsubsection{SUSY Extension}

The supersymmetric model is constructed adding scalar fields in the complex
representation in order to avoid chiral anomalies from their superpartners.
With this spectrum we obtain the following values for the $b_{i}$
coefficients 
\begin{eqnarray}
(b_{Y},b_{2L},b_{s}) &=&\left( 28,6,-3\right) ,  \notag  \label{betaYESUSY}
\\
(b_{X},b_{3L},b_{3C}) &=&\left( 42,5,0\right) .  \label{betaXESUSY}
\end{eqnarray}%
We also consider an extended spectrum for the scalar sector. With these
values we get only an upper limit for $M_{X}$ in the \textbf{ZSBS} 
\begin{equation}
M_{X}\leq 2.18\times 10^{8}{\text{G}eV}.
\end{equation}%
But the restriction in the normalization parameter $a$ in Eq. %
\eqref{pararestric}, requires $M_{X}\geq 8.1\times 10^{8}$. This situation
discards the model with only one intermediate breaking.

For the \textbf{XSBS}, the hierarchy condition gives only a lower limit $%
6.87\times 10^{13}$GeV$\leq M_{X}$. But $a^{2}\geq 3$ gives $M_{X}\geq
1.35\times 10^{22}GeV$, discarding the model.

\subsection{Model F}

It is also possible to obtain cancellation of anomalies by a tiny variation
of the fermionic spectrum from the model D of Pleitez-Tonasse \cite{Pleitez3}%
. This new spectrum will be given by 
\begin{eqnarray}
\psi _{L}^{\alpha } &=&(\nu ^{\alpha },e^{\alpha -},E^{{\alpha }%
+})_{l}^{T}\sim (1,3,0),  \notag \\
e_{L}^{\alpha +} &\sim &(1,1,-1),\;\;\;\;E_{L}^{\alpha +}\sim (1,1,1).
\end{eqnarray}%
where an exotic lepton has been introduced in the third entry of the triplet
instead of the electron right-handed part, such that there are also leptonic
singlets of $SU(3)_{L}$.

The $b_{i}$ coefficients at low energy scales differ from those for the
model E in the contribution due to the sextuplet. Therefore the values are 
\begin{eqnarray}
(b_{Y},b_{2L},b_{s}) &=&\left( 26/3,-3,-7\right) ,  \notag  \label{betaYF} \\
(b_{X},b_{3L},b_{3C}) &=&\left( 26,-\frac{13}{2},-5\right) .
\end{eqnarray}%
With these coefficients, the allowed region for $M_{X}$ is given by 
\begin{equation}
2.05\times 10^{16}GeV\leq M_{X}\leq 3.15\times 10^{16}GeV.
\end{equation}%
The restriction from Eq. (\ref{pararestric}) concerning the parameter $a^{2}$%
, gives an upper bound of $M_{X}\leq 4.1\times 10^{17}GeV$, i.e. no
additional restriction.

\subsubsection{SUSY Extension}

The extension is done analogously to the model E, obtaining the following
values for the $b_{i}$ coefficients 
\begin{eqnarray}
(b_{Y},b_{2L},b_{s}) &=&\left( 22,2,-3\right) ,  \notag \\
(b_{X},b_{3L},b_{3C}) &=&\left( 48,0,0\right) .
\end{eqnarray}%
The SUSY version is in the \textbf{2UP} for both breaking schemes. In the 
\textbf{ZSBS}, $M_{X}$ reads%
\begin{equation}
M_{X}=2.76\times 10^{13}GeV.
\end{equation}%
In this case, the condition $a^{2}\geq 1/3$ imposes a strong bound on the $%
M_{U}$ scale. From

\begin{equation*}
2.76\times 10^{13}\text{GeV}\leq M_{U}\leq 1.13\times 10^{15}\text{GeV}
\end{equation*}%
we get $a^{2}=\left[ 1/3,\ 1.67\right] $. Larger values of $M_{U}$ leads to
non allowed values of $a^{2}$ (less than $1/3$).

In the \textbf{XSBS} we have $M_{X}=2.05\times 10^{16}GeV$. Using $M_{U}\in %
\left[ 2\times 10^{16},\ 10^{17}\right] $ GeV we get $a^{2}\in \left[ 1.55,\
1.82\right] $. They satisfy the condition $a^{2}\geq 1/3$. Hence, SUSY
versions give a possible scenario of unification in our scheme.

\section{Some possible scenarios with groups of grand unification}

One of our main assumptions was that there is no necessarily a group of
grand unification at the scale in which the coupling constants converge. In
this way, the $a^{2}$ parameter becomes free, and indeed we could reverse
the problem in a certain way, since we have in many cases an allowed region
for $a^{2}$ and we could ask what groups of grand unification (if any) could
lead to values of $a^{2}$ admitted by our scheme. Let us elaborate about
this possibility

After working some 331 models with one or three families, we can see that in
some of them is possible to find a scale $M_{X}$ and a normalization factor $%
a^{2}$ that gives unification of the coupling constants. In the absence of a
group of grand unification (GUT) $a^{2}$ is not fixed by the group
structure. Notwithstanding, UCC imposes some restrictions on $a^{2}$, and by
using these values we can look for simple groups containing a 331 group and
fixing an $a^{2}$ belonging to the allowed interval mentioned above. Some
good examples of GUT$\ $are$\ E_{6}$, and $SU\left( 7\right) $.

The group $E_{6}$ can be broken into $SU\left( 3\right) _{C}\otimes SU\left(
3\right) _{L}\otimes SU\left( 3\right) _{R}$ or $SU\left( 6\right)
_{L}\otimes SU\left( 2\right) $. Models $A$ and $B$ shown in this paper
could be embedded in this scheme. On the other hand, the 331 models of three
families can be embedded into $SU\left( 7\right) $, thus, we shall study the
latter scenario in more detail.

There are different versions of the $SU(7)$ GUT that lead to 331 models
according to the irreducible representations (irrep) that cancel anomalies,
and the definition of the electric charge or equivalently the linear
combination of diagonal generators that define the hypercharge\ (i.e. the
normalizing factor $a^{2}$). The combination of irreducible representations
free of anomalies, permits to accomodate those 331 models with three
families by the branching rules. The additional singlets or exotic fields
acquire masses over the $M_{X}$ scale, and do not affect the RGE. There are
different combinations of anomaly free irreps. Taking into account the
irreps $\Psi _{(1)}^{\alpha }[7]$, $\Psi _{(3)}^{\alpha \beta }[21]$ and $%
\Psi _{(2)}^{\alpha \beta \gamma }[35]$, where the subindex means the
anomaly coefficient, the bracket coefficient means the dimension and the
labels $\alpha ,\beta ,\gamma =1,\cdots ,7$. Models can be classified if no
irrep appears more than once i.e. a form like $\Psi _{(-1)\alpha }\oplus
\Psi _{(3)}^{\alpha \beta }\oplus \Psi _{(-2)\alpha \beta \gamma }$ ; but in
general the same irrep can be repeated such as $5\times \Psi _{(-1)\alpha
}\oplus \Psi _{(3)}^{\alpha \beta }\oplus \Psi _{(2)}^{\alpha \beta \gamma }$%
. On the other hand, the branching rules according to $(SU(3)_{c},SU(2)_{L})$
are given by 
\begin{eqnarray}
\Psi ^{\alpha }\oplus \Psi ^{\alpha \beta }\oplus \Psi ^{\alpha \beta \gamma
} &=&\left[ (3^{\ast },1)+(1,2)+(1,1)+(1,1)\right] _{7}\oplus \left[
(3^{\ast },1)+(3,2)\right.  \notag \\
&+&\left. (3,1)+(3,1)+(1,1)+(1,2)+(1,2)+(1,1)\right] _{21}  \notag \\
&\oplus &\left[ (1,1)+(3,2)+(3,1)+(3,1)+(3^{\ast },1)+(3^{\ast },2)\right. 
\notag \\
&+&\left. (3^{\ast },2)+(3^{\ast },1)+(1,1)+(1,1)+(1,2)\right] _{35}
\label{descompo}
\end{eqnarray}%
The electromagnetic charge of the particles can be chosen by defining the
hypercharge which is a linear combination of the $U(1)$ factors of the $%
SU(7) $, and imposing the condition that the singlet of color have
electromagnetic charges $q=\pm 1/3,\pm 2/3,\pm 4/3,\pm 5/3\ $\footnote{%
The values $4/3,5/3$ correspond to exotic quark charges that arise in some
331 models, like model E.} or that the electromagnetic charge of the
leptonic singlets are $\pm 1$, $0$.

Assuming an unification scheme with the simple group $SU(7)$ and passing
through a 331 model with three families, by one step of symmetry breaking,
we get the scheme 
\begin{eqnarray}
SU(7)\overset{M_{U}}{\rightarrow } &&SU(3)_{c}\otimes SU(3)_{L}\otimes
U(1)_{X}  \notag \\
\overset{M_{X}}{\rightarrow } &&SU(3)_{c}\otimes SU(2)_{L}\otimes U(1)_{Y}
\label{scheme}
\end{eqnarray}%
the assignment of the electromagnetic charge is of the form%
\begin{eqnarray*}
Q &=&T_{3L}+\alpha Y^{a}+\beta Y^{b}+\gamma Y^{c}\equiv T_{3L}+a\widetilde{Y}
\\
a^{2} &\equiv &\alpha ^{2}+\beta ^{2}+\gamma ^{2}
\end{eqnarray*}%
where $Y^{a},\ Y^{b},\ Y^{c}$ have the same normalization as the $T_{3L}$, $%
\widetilde{Y}$ generators, and they correspond to the $U\left( 1\right) $
abelian subgroups induced from different subalgebras of $SU(7)\supset
SU\left( n\right) _{L}\otimes SU\left( m\right) _{C}\otimes U(1)^{a}$, where 
$SU\left( n\right) _{L}\supset SU(2)_{L}\otimes U(1)^{b}$ and $SU\left(
m\right) _{C}\supset SU(3)_{c}\otimes U(1)^{c}$. If the fundamental
representation is decomposed as in Eq. (\ref{descompo}), the most general
assignment is 
\begin{equation}
Q=diag(q,q,q,b,a,a-1,1-3q-2a-b).  \label{carga}
\end{equation}%
And we can identify the charge of $(3,1)$ to be $-1/3$ and those of $(1,2)$
as $(1,0)$, then 
\begin{equation}
Q=diag(-1/3,-1/3,-1/3,b,1,0,-b)  \label{carga1}
\end{equation}%
where $b=0,\pm 1$ if we have singlet leptons or $b=\pm 1/3,\pm 2/3,\pm
4/3,\pm 5/3$ if we have singlet quarks.

From the following possible $SU\left( 7\right) $ maximal subalgebras 
\begin{eqnarray*}
Model\ I &&\ SU\left( 4\right) _{C}\otimes SU\left( 3\right) _{L}\otimes
U\left( 1\right) ^{a} \\
Model\ II &&\ \ SU\left( 3\right) _{C}\otimes SU\left( 4\right) _{L}\otimes
U\left( 1\right) ^{a} \\
Model\ III &&SU(6)_{L}\otimes U(1)^{a}\rightarrow SU\left( 3\right)
_{C}\otimes SU\left( 3\right) _{L}\otimes U\left( 1\right) ^{c}\otimes
U\left( 1\right) ^{a}
\end{eqnarray*}%
we can settle three different assignments for the hypercharge according to
the scheme in Eqs. (\ref{carga}, \ref{carga1}). For example, Model I is
known in the literature as the Pati Salam model. The generators are defined
by 
\begin{eqnarray}
Y^{a} &=&\sqrt{\frac{6}{7}}diag\left( 1/4,1/4,1/4,1/4,-1/3,-1/3,-1/3\right) 
\notag \\
Y^{b} &=&\sqrt{\frac{1}{3}}diag\left( 0,0,0,0,1/2,1/2,-1\right)  \notag \\
Y^{c} &=&\sqrt{\frac{3}{8}}diag\left( 1/3,1/3,1/3,-1,0,0,0\right)
\end{eqnarray}%
In this case $Y^{b}$ corresponds to the generator $T_{8}$ of $SU(3)_{L}$;
and a linear combination of $Y^{a}$, $Y^{c}$ corresponds to $U(1)_{X}$ of
the 331 models. For model II the generators are defined by 
\begin{eqnarray}
Y^{a} &=&\sqrt{\frac{6}{7}}diag\left( 1/3,1/3,1/3,-1/4,-1/4,-1/4,-1/4\right)
\notag \\
Y^{b} &=&\sqrt{\frac{1}{3}}diag\left( 0,0,0,0,1/2,1/2,-1\right)  \notag \\
Y^{c} &=&\sqrt{\frac{3}{8}}diag\left( 0,0,0,-1,1/3,1/3,1/3\right)
\end{eqnarray}%
And for model III the generators are defined by 
\begin{eqnarray}
Y^{a} &=&\sqrt{\frac{3}{7}}diag\left( 1/6,1/6,1/6,-1,1/6,1/6,1/6\right) 
\notag \\
Y^{b} &=&\sqrt{\frac{1}{12}}diag\left( 1,1,1,0,-1,-1,-1\right)  \notag \\
Y^{c} &=&\sqrt{\frac{1}{3}}diag\left( 0,0,0,0,1/2,1/2,-1\right)
\end{eqnarray}%
By choosing the values for the free parameter $b=0,\pm 1$ for leptons, and $%
b=\pm 1/3,\pm 2/3,\pm 4/3,\pm 5/3$ for quarks, the coefficients $\alpha $, $%
\beta $, $\gamma \ $that determine the hypercharge can be obtained, and they
are displayed in table \ref{tab:hyper}. On the other hand, table \ref{tab:a2}
shows the $a^{2}$ normalization factor of the hypercharge according to the $%
b $ factor.

The next step is to check whether the three family models consider here
(models C, D, E, F) can be accomodated properly in a $SU\left( 7\right) $
GUT. It can be done by comparing the allowed region for $a^{2}$ in table \ref%
{tab:a2permitido}, with the values of $a^{2}$ in table \ref{tab:a2}. The
second column shows the allowed interval of $a^{2}$ obtained in previous
sections in the absence of a grand unification group, the third column shows
the allowed interval by taking into account additional restrictions coming
from proton decay ($M_{U}\gtrsim 2\times 10^{16}$GeV), that arise when a GUT
theory is introduced\footnote{%
There is another correction coming from the new spectrum introduced, but we
have assumed that the new heavy modes do not alter the RGE significantly.}.

The restrictions displayed in table \ref{tab:a2permitido} for non SUSY
versions of models C, D, F; show that they can be embedded in $SU\left(
7\right) $ since the bound $a^{2}\geq 1/3$ is accomplished by all the values
of $a^{2}$ in table \ref{tab:a2}. Model E (the only one with exotic charges
studied here) has been ruled out from phenomenological grounds under our
scheme. On the other hand, from table \ref{tab:a2permitido}, we see that the 
$a^{2}$ parameter is strongly restricted in the SUSY versions of these
models, and not all of them can be accomodated in $SU\left( 7\right) $
according to table \ref{tab:a2}.

\begin{table}[tbp]
\begin{center}
\begin{tabular}{||c||ccc||ccc||ccc||}
\hline\hline
b &  & Model I &  &  & Model II &  &  & Model III &  \\ \hline\hline
& $\sqrt{\frac{7}{6}}\alpha$ & $\sqrt{3}\beta$ & $\sqrt{\frac{8}{3}}\gamma$
& $\sqrt{\frac{7}{6}}\alpha$ & $\sqrt{3}\beta$ & $\sqrt{\frac{8}{3}}\gamma$
& $\sqrt{\frac{7}{3}}\alpha$ & $\sqrt{\frac{1}{12}}\beta$ & $\sqrt{\frac{1}{3%
}}\gamma$ \\ \hline\hline
$0$ & $-1$ & $+\frac{1}{3}$ & $-\frac{1}{4}$ & $-1$ & $+\frac{1}{3}$ & $+%
\frac{1}{4}$ & $0$ & $-\frac{1}{3}$ & $+\frac{1}{3}$ \\ \hline
$1$ & $0$ & $1$ & $-1$ & $-1$ & $1$ & $-\frac{3}{4}$ & $-1$ & $-\frac{1}{6}$
& $1$ \\ \hline
$-$1 & $-2$ & $-\frac{1}{3}$ & $+\frac{1}{2}$ & $-1$ & $-\frac{1}{3}$ & $+%
\frac{5}{4}$ & $1$ & $-\frac{1}{2}$ & $-\frac{1}{3}$ \\ \hline
$+\frac{1}{3} $ & $-\frac{2}{3}$ & $+\frac{5}{9}$ & $-\frac{1}{2}$ & $-1$ & $%
+\frac{5}{9}$ & $-\frac{1}{12}$ & $-\frac{1}{3}$ & $-\frac{5}{18}$ & $+\frac{%
5}{9}$ \\ \hline
$-\frac{1}{3}$ & $-\frac{4}{3}$ & $-\frac{1}{9}$ & $0$ & $-1$ & $+\frac{1}{9}
$ & $+\frac{7}{12}$ & $+\frac{1}{3}$ & $-\frac{7}{18}$ & $+\frac{1}{9}$ \\ 
\hline
$+\frac{2}{3} $ & $-\frac{1}{3}$ & $+\frac{7}{9}$ & $-\frac{3}{4}$ & $-1$ & $%
+\frac{7}{9}$ & $-\frac{5}{12}$ & $-\frac{2}{3}$ & $+\frac{2}{9}$ & $+\frac{7%
}{9}$ \\ \hline
$-\frac{2}{3} $ & $-\frac{5}{3}$ & $-\frac{1}{9}$ & $+\frac{1}{4}$ & $-1$ & $%
-\frac{1}{9}$ & $+\frac{11}{12}$ & $+\frac{2}{3}$ & $-\frac{4}{9}$ & $-\frac{%
1}{9}$ \\ \hline
$+\frac{4}{3} $ & $+\frac{1}{3}$ & $+\frac{11}{9}$ & $-\frac{3}{2}$ & $-1$ & 
$+ \frac{11}{9}$ & $-\frac{13}{12}$ & $-\frac{4}{3}$ & $-\frac{1}{9}$ & $+%
\frac{11}{9}$ \\ \hline
$-\frac{4}{3} $ & $-\frac{7}{3}$ & $-\frac{5}{9}$ & $+\frac{3}{4}$ & $-1$ & $%
-\frac{5}{9}$ & $+\frac{19}{12}$ & $+\frac{4}{3}$ & $-\frac{5}{9}$ & $-\frac{%
5}{9}$ \\ \hline
$+\frac{5}{3} $ & $+\frac{2}{3}$ & $+\frac{13}{9}$ & $-\frac{3}{2}$ & $-1$ & 
$+ \frac{13}{9}$ & $-\frac{17}{12}$ & $-\frac{5}{3}$ & $-\frac{1}{18}$ & $+%
\frac{13}{9}$ \\ \hline
$-\frac{5}{3} $ & $-\frac{8}{3}$ & $-\frac{7}{9}$ & $1$ & $-1$ & $-\frac{7}{9%
}$ & $+\frac{23}{12}$ & $+\frac{5}{3}$ & $-\frac{11}{18}$ & $-\frac{7}{9}$
\\ \hline\hline
\end{tabular}%
\vspace{.5cm}
\end{center}
\caption{Hypercharge definition using the embbeding of $U(1)$ into
subalgebras of $SU(7)$ for models I, II, III}
\label{tab:hyper}
\end{table}

\begin{table}[tbp]
\begin{center}
\begin{tabular}{||c||c||}
\hline\hline
b & $a^{2}$ \\ \hline\hline
$0$ & $\frac{5}{3}\simeq 1.\,\allowbreak 6\,7$ \\ \hline
$\pm 1$ & $\frac{17}{3}\simeq 5.\,\allowbreak 6\,7$ \\ \hline
$\pm \frac{1}{3}$ & $\frac{19}{9}\simeq 2.\,\allowbreak 11$ \\ \hline
$\pm \frac{2}{3}$ & $\frac{31}{9}\simeq 3.\,\allowbreak 44$ \\ \hline
$\pm \frac{4}{3}$ & $\frac{79}{9}\simeq 8.\,\allowbreak 7\,8$ \\ \hline
$\pm \frac{5}{3}$ & $\frac{115}{9}\simeq 12.\,\allowbreak 78$ \\ \hline\hline
\end{tabular}
\vspace{0.5cm}
\end{center}
\caption{Hypercharge definition using the embbeding of $U(1)$ into
subalgebras of $SU(7)$ for models I, II, III}
\label{tab:a2}
\end{table}
\begin{table}[tbp]
\begin{center}
\begin{tabular}{||l||l||l||}
\hline\hline
& 
\begin{tabular}{l}
Allowed interval for $a^{2}\ $with no \\ 
restriction from proton decay%
\end{tabular}
& 
\begin{tabular}{l}
Allowed interval for $a^{2}\ $with \\ 
restriction from proton decay%
\end{tabular}
\\ \hline\hline
non SUSY C, D, F & $\geq 1/3$ & $\geq 1/3$ \\ \hline\hline
C\ (SUSY ZSBS) & $\left[ 1.78,\ 3.26\right] $ & $\left[ 2.07,\ 3.26\right] $
\\ \hline\hline
C\ (SUSY XSBS) & $\left[ 1.84,\ 1.97\right] $ & $\left[ 1.84,\ 1.97\right] $
\\ \hline\hline
D\ (SUSY ZSBS) & $\left[ 2.27,\ 3.26\right] $ & $\left[ 2.47,\ 3.26\right] $
\\ \hline\hline
D\ (SUSY XSBS) & $\left[ 1.88,\ 1.97\right] $ & $\left[ 1.88,\ 1.97\right] $
\\ \hline\hline
F\ (SUSY ZSBS) & $\left[ 1/3,\ 1.67\right] $ & Excluded \\ \hline\hline
F\ (SUSY XSBS) & $\left[ 1.55,\ 1.82\right] $ & $\left[ 1.55,\ 1.82\right] $
\\ \hline\hline
\end{tabular}%
\end{center}
\caption{Allowed interval for the normalization $a^{2}$ for models with
three families. In the second column, no restrictions from proton decay are
taken into account, while in the third column they are. Models E (SUSY and
non SUSY) were discarded from phenomenological grounds.}
\label{tab:a2permitido}
\end{table}

\section{Conclusions}

\begin{table}[tbh]
\begin{center}
\setlength{\belowcaptionskip}{2pt} 
\scalebox{0.93}[0.83]{\rotatebox{0}{\mbox{\renewcommand{\arraystretch}{1.7}
\begin{tabular}{||c||c|c|c||}
\hline\hline
& \multicolumn{3}{|c|}{Allowed interval for $M_{X}$} \\ \hline\hline
Model & No SUSY & NSBS & SBS \\ \hline
A & $\left[ 1.63\times 10^{16},\ 2.05\times 10^{16}\right] $ & $\left[
1.27\times 10^{8}{,\ }2.76\times 10^{13}\right] $ & $\left[ 1.76\times
10^{14}{,\ }2.05\times 10^{16}\right] $ \\ \hline
B & $\left[ 1.63\times 10^{16}{\text{,\ }}2.05\times 10^{16}\right] $ & No
allowed & $\left[ 1.76\times 10^{14}{,\ }2.05\times 10^{16}\right] $ \\ 
\hline
C & $\left[ 2.05\times 10^{16},\ 3.15\times 10^{16}\right] $ & $=2.76\times
10^{13}$ & $=2.05\times 10^{16}$ \\ \hline
D & $\left[ 2.05\times 10^{16},\ 3.15\times 10^{16}\right] $ & $=2.76\times
10^{13}$ & $=2.05\times 10^{16}$ \\ \hline
E & No allowed & No allowed & No allowed \\ \hline
F & $\left[ 2.05\times 10^{16},\ 3.15\times 10^{16}\right] $ & $=2.76\times
10^{13}$ & $=2.05\times 10^{16}$ \\ \hline\hline
\end{tabular}}}}
\end{center}
\par
\vspace{.5cm}
\caption{Summary of the results obtained in the paper for the six different
331 models studied, as well as their corresponding SUSY extensions}
\label{tab:summary}
\end{table}

We have studied the possibility of having a 331 model as the only
intermediate group between the electroweak scale and the scale $M_{U}$ in
which the three coupling constants unify. We assume that there is no
necessarily a simple gauge group at the $M_{U}$ scale. From the analysis of
the renormalization group equations (RGE) we examine different 331 models of
one and three families as well as their supersymmetric extensions.

Specifically, we are supposing that the three different couplings unify at
certain scale $M_{U}$, and a symmetry breaking to a gauge group $%
SU(3)_{L}\otimes U(1)_{X}$ occurs at a lower\ scale $M_{X}$. Then, a second
breaking occurs at the $M_{Z}$ scale to arrive at the SM gauge group. We are
also assuming that all particles beyond the SM are getting masses of the
order of $M_{X}$. Other analyses can be done by supposing that the particle
contents are in different thresholds. For our analysis of the RGE, the
following conditions are taking into account: \ding{182} The hierarchy
condition given by Eq. (\ref{gerarquia}) i.e. $M_{X}\leq M_{U}\leq 1\times
10^{17}$ GeV, should be satisfied. \ding{183} The condition described by Eq.
(\ref{pararestric}) should be fulfilled, where $a\ $is the normalization
parameter of the hypercharge $Y$, that leads to the unification of the
couplings at certain scale. \ding{184} There is no neccesarily a grand
unification gauge group at the $M_{U}$ scale.

In the case of the supersymmetric extensions, the RGE analysis depends on
the specific supersymmetric version but also on the supersymmetry breaking
scale. In particular, we assume two possible SUSY breaking scenarios: SUSY
breaking at electroweak scale and SUSY breaking at the scale of 331
breaking. Although SUSY breaking at electroweak scale is not a realistic
scenario, numerical results do not change significantly with respect to the
more realistic framework with SUSY breaking at some few TeV's.

Based on the criteria explained above, we can either find an allowed
interval for the 331 breaking scale $M_{X}$ or rule out the model as a
possible grand unified theory under the scheme described above. We summarize
the results found in this paper in table \ref{tab:summary} for the six
different 331 models studied here as well as their SUSY extensions (some
SUSY extensions that are ruled out have not been included). In many cases,
we see that SUSY versions tend to give lower allowed values for the 331
breaking scales than their non SUSY counterparts, it could make SUSY
extensions easier to test from the phenomenological point of view. It worths
emphasizing that taking into account that the spectrum out of the SM lies at
the $M_{X}$ scale, when assuming that the SUSY breaking occurs at the $M_{X}$
scale we are in a natural scenario for split Supersymmetry \cite{Romanino}-\cite{Romanino2}.

On the other hand, we see that models $C$ and $D$ predicts the same allowed
intervals for $M_{X}$ in SUSY and non SUSY versions. However, SUSY versions
predict different values for the normalization parameter $a^{2}$\textbf{.}

In addition, by finding the allowed values for $a^{2}$ we can proceed to see
what groups of grand unification could give a value of $a^{2}$ lying in the
allowed range. Although the introduction of a group at $M_{U}$ scale
introduces new singlets that could lead to tiny changes in the RGE analysis
from $M_{X}$ to $M_{U}$, this procedure helps us to figure out what
scenarios of grand unification (if any) are possible under our scheme. In
particular, we find that some 331 models with three families, can be
properly embedded in a grand unification scenario with $SU\left( 7\right) $,
especially in non susy frameworks.

As a matter of perspectives, two loops analysis for RGE can be carried out 
\cite{MartinezM} in order to fit the $M_{U}$ scale better, especially taking
into account the uncertainty in the starting coupling constants at
electroweak scale. However, significant changes in the allowed regions
obtained here are not expected for a well behaved perturbative regime. On
the other hand, it is possible to unify the coupling constant associated to $%
U\left( 1\right) $ at string scale instead of the $M_{U}$ scale \cite{25}-\cite{252}.

Finally, it worths saying that the non allowed models are not neccesarily
ruled out. We only can say that they cannot produce \textbf{UCC} under the
scheme in which 331 is the only gauge symmetry between the $M_{U}$ and SM
scales. For instance, it could be possible that they achieve unification by
either introducing new physics, or requiring extra breaking steps from the $%
M_{U}$ scale to SM scale. By introducing new physics, it is possible to get
a lower $M_{X}$ scale of the order of TeV accesible to the LHC.

\section{Acknowledgements}

We thank Colciencias, Fundación Banco de la República, and High Energy Latin
American European Network for its financial support.

\end{document}